\documentclass[fleqn,twoside]{article}
\usepackage{espcrc2,epsfig,latexsym}

\newcommand{\Mpi}{M_\pi}
\newcommand{\mpi}{m_\pi}
\newcommand{\Fpi}{F_\pi}
\newcommand{\Mrho}{M_\rho}
\newcommand{\mrho}{m_\rho}

\title{Volume dependence of light hadron masses in full lattice QCD} 

\author{B.~Orth\address{Department of Physics, University of
  Wuppertal, Gau{\ss}str.~20, D-42097 Wuppertal, Germany},
  T.~Lippert\addressmark\ and K.~Schilling\addressmark}

\begin{document}

\begin{abstract}
  The aim of the GRAL project is to simulate full QCD with standard
  Wilson fermions at light quark masses on small to medium-sized
  lattices and to obtain infinite-volume results by extrapolation. In
  order to establish the functional form of the volume dependence we
  study systematically the finite-size effects in the light hadron
  spectrum.  We give an update on the status of the GRAL project and
  show that our simulation data for the light hadron masses depend
  \emph{exponentially} on the lattice size.
\end{abstract}

\maketitle

\section{INTRODUCTION}

Lattice QCD has always been plagued by the fact that towards realistic
quark masses the simulation costs increase drastically due to the
large correlation lengths of the light states and the large lattice
volumes that one employs in order to avoid finite-size (FS) effects.
This difficulty has, in fact, prevented us until this day from
carrying out simulations with truly physical parameters.  Instead one
resorts to computations with several un-physically large quark masses
in FS-effect-free volumes, followed by an extrapolation to the
physical quark mass.  However, the chiral extrapolation introduces
uncertainties that can be reduced only by increasing the overlap of
lattice simulations with ChPT.  It is thus inevitable to look out for
feasible ways to obtain lattice results at light quark masses.
Current approaches like overlap, domain-wall or improved staggered
fermions are still very costly, especially if one goes beyond the
quenched approximation. Faced with limited computer resources we
therefore investigate the feasibility of an approach that restricts
itself to the use of standard Wilson fermions.

The aim of the \emph{GRAL} project is to carry out a systematic study
of FS effects in the light hadron spectrum in the parameter regime
accessible to us. With simulations on series of small to medium-sized
lattice volumes (at fixed coupling and quark mass, respectively) we
also want to address the issue of whether one can use formulae
obtained from this investigation to extrapolate hadronic observables
to the infinite volume.

\section{VOLUME DEPENDENCE OF THE LIGHT HADRON MASSES}

In 1983 M.~L\"uscher published a universal field theoretic formula for
the shift in the mass of a stable particle enclosed in a finite
box~\cite{Luscher:1983rk,Luscher:1985dn}. The formula states for
asymptotically large volumes that the FS mass-shift vanishes
exponentially with increasing box size at a rate that depends on the
particle considered and on the spectrum of light particles in the
theory. If, in QCD, we consider a stable hadron in a box of size $L^3$
with periodic boundary conditions and sufficiently large (Euclidian)
time-extent $T$, its mass $M_H(L)$ becomes a universal function of
$M_\pi L$ (where, at fixed $\beta$, $M_\pi=\lim_{L\to\infty}
M_\pi(L)$) in the finite-volume continuum limit. Since FS effects
probe the system at large distances and are thus insensitive to the
form and magnitude of the UV cutoff, this function is expected to hold
also for finite lattice spacings. With the (constant) LO chiral
expression for the $\pi$-$\pi$ forward scattering amplitude the
relative mass-shift for the pion is given by
\begin{eqnarray}
\label{eqn:pion}
\frac{\Mpi(L)-\Mpi}{\Mpi} &=& \frac{3}{8\pi^2} \frac{\Mpi^2}{\Fpi^2}
\frac{K_1(\Mpi L)}{\Mpi L} \\
& \sim &  \frac{3}{4(2\pi)^{3/2}} \frac{\Mpi^2}{\Fpi^2}
\frac{e^{-\Mpi L}}{(\Mpi L)^{3/2}}
\end{eqnarray}

If we assume, for simplicity, the pion-nucleon forward scattering
amplitude to be constant, too, then the formula for the nucleon
involves a term that is---up to low-energy constants---proportional to
the one for the pion. In addition it contains a term proportional to
\begin{equation}
\label{eqn:nucl}
\exp\left(-\Mpi L\sqrt{1-\Mpi^2/(4 M_N^2)}\right)/(\Mpi L).
\end{equation}

On the other hand, in 1992 Fukugita \emph{et al.} found a power-like
$L$-dependence of the form
\begin{equation}
\label{eqn:l3}
M_H(L) = M_H + c/L^3
\end{equation}
in their data for the pion, rho and nucleon
masses~\cite{Fukugita:1992jj,Fukugita:1992hr}. The usual explanation
for the apparent discrepancy is that L\"uscher's formula deals with
asymptotically large lattice volumes where FS effects arise from a
squeezing of the virtual pion cloud surrounding the hadron in a box
with periodic boundary conditions. In contrast, Fukugita \emph{et
  al.}\ ascribe the power-like behaviour of their data to a squeezing
of the hadron itself (as it would occur for small lattice volumes). In
this picture one expects a power-like dependence for sub-asymptotic
lattice volumes that gradually changes into an exponential fall-off of
the hadron masses towards larger box sizes.

\section{GRAL STATUS}

\begin{table*}[htb]
\caption{GRAL production status as of 07/08/2003}
\label{tab:production_status}
\newcommand{\cc}[1]{\multicolumn{1}{c}{#1}}
\renewcommand{\arraycolsep}{0.5pc} 
\renewcommand{\arraystretch}{1.2} 
$
 \begin{array}{@{}llllllllll}
  \hline
  \cc{\beta} & \cc{\kappa} & \cc{L^3 \times T} & \cc{N_\mathrm{conf}} &
  \cc{\langle \frac{1}{3} \mathrm{Tr}\,\Box \rangle} & \cc{r_0} &
  \cc{a_{r_0}^{-1}\;[\mathrm{GeV}]} & \cc{\Mpi(L)\,L} & \cc{\Mpi L} & \cc{L\;[\mathrm{fm}]} \\
  \hline
 5.32 & 0.1665          & 12^3 \times 32 & 4900 & 0.5395(4)  &         &         &         &         &          \\
      &                 & 14^3 \times 32 & 2300 & 0.5382(2)  &         &         &         &         &          \\
      &                 & 16^3 \times 32 & 3400 & 0.5382(1)  & 3.84(3) & 1.51(2) & 4.3(2)  &         &          \\
 \hline
 5.6  & 0.1575          & 10^3 \times 32 & 6200 & 0.57324(7) &         &         & 4.9(2)  & 2.76(3) & 0.850(4) \\
      &                 & 12^3 \times 32 & 7200 & 0.57280(5) &         &         & 4.41(8) & 3.32(4) & 1.020(5) \\
      &                 & 14^3 \times 32 & 6500 & 0.57262(4) &         &         & 4.23(7) & 3.87(4) & 1.190(6) \\
      & \mbox{SESAM}    & 16^3 \times 32 & 6500 & 0.57254(3) & 5.96(8) & 2.35(3) & 4.49(6) & 4.43(5) & 1.360(7) \\
      & \mbox{T$\chi$L} & 24^3 \times 40 & 5100 & 0.57248(1) & 5.89(3) & 2.32(1) & 6.64(7) & 6.64(7) & 2.04(1)  \\
 \cline{2-10}
      & 0.1580          & 14^3 \times 32 & 4400 & 0.57367(4) &         &         & 4.0(2)  &         &          \\
      &                 & 16^3 \times 32 & 3900 & 0.57345(5) &         &         & 3.7(2)  &         &          \\
      & \mbox{T$\chi$L} & 24^3 \times 40 & 4500 & 0.57337(2) & 6.23(6) & 2.45(2) & 4.78(7) &         &          \\
 \hline
 \end{array}
$
\end{table*}

Table~\ref{tab:production_status} summarises some details of our
simulations. We use the standard unimproved Wilson action with two
degenerate flavours of dynamical fermions. Our HMC codes are running
on the APEmille at DESY/Zeuthen and on the cluster installation ALiCE
at Wuppertal University. In addition to the already available SESAM
($L=16$) and T$\chi$L ($L=24$) ensembles we are generating
configurations on smaller volumes with $L$ varying between 10 and 16.
The figures referring to these runs are, where available, preliminary.
With our simulations at $\beta=5.32$, $\kappa=0.1665$ we aim at a
quark mass corresponding to $\Mpi/\Mrho\approx 0.5$, \textit{i.e.}\ 
below the smallest $\Mpi/\Mrho$ of 0.575(16) previously achieved by
the SESAM/T$\chi$L project. In this article we focus on the ensembles
at $\beta=5.6$, $\kappa=0.1575$ where we assume the largest,
$24^3\times 40$-lattice to be free of FS effects.

\section{RESULTS}

\begin{figure}[htb]
 \begin{center}
  \epsfig{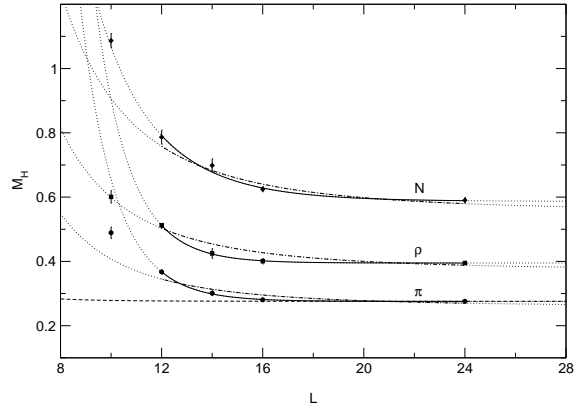}
  \caption{\label{fig:fss} Dependence of the $\pi$-, $\rho$- and
    $N$-masses for $\beta=5.6$, $\kappa=0.1575$ on the spatial lattice
    extent, $L$. The curves are explained in the text.}
 \end{center}
\end{figure}
The (preliminary) masses we obtain for the pion, rho and nucleon are
displayed in Fig.~\ref{fig:fss}.

The flat dashed curve represents L\"uscher's formula for the pion
mass, Eq.~(\ref{eqn:pion}), with $\Mpi$ and $\Fpi$ as obtained from
the $24^3$-lattice. From this curve it is obvious that the data points
at $L=10,12,14$ are outside the chiral regime. This can also be seen
from $\Fpi L$, which takes the values 0.91(3), 1.10(3), 1.28(4),
1.46(4) and 2.19(6) for $L=10,12,14,16,24$, respectively. Since ChPT
is a low-energy (large distance) expansion it is only valid for $\Fpi
L\gg 1$ and $\Mpi/(4\pi\Fpi)\ll 1$; $\Mpi/(4\pi \Fpi)$ is 0.240(7) for
$L=24$.  Nevertheless we observe that the pion data can be well fitted
from $L=12$ with the ``semi-empirical'' formula
\begin{equation}
\Mpi(L) = \mpi + a_1 \frac{3}{4(2\pi)^{3/2}} \frac{\mpi^{3/2}}{\Fpi^2}
\frac{e^{-a_2\mpi L}}{L^{3/2}}
\end{equation}
where $\mpi$, $a_1$ and $a_2$ are free fit-parameters. $a_1$ and $a_2$
are chosen such that their deviation from 1 indicates ``how far away''
we are from the original formula~(\ref{eqn:pion}). For the rho (which
cannot decay on the employed lattices) and the nucleon we adopt, for
various practical reasons, the generic ansatz
\begin{equation}
M_{\rho,N}(L) = m_{\rho,N} + a_1 \frac{e^{-a_2\mpi L}}{L^{3/2}},
\end{equation}
\textit{i.e.}\ we effectively drop the term proportional
to~(\ref{eqn:nucl}). Here, $\mpi$ is fixed to the value obtained from
the pion fit. All the fits are displayed in Fig.~\ref{fig:fss}, where
the lines are solid within the fit range and dotted outside.  For
comparison the dash-dotted curves show $1/L^3$-fits of the
form~(\ref{eqn:l3}). The $\chi^2/\mathrm{dof}$-values for the
exponential fits are 0.20, 0.03 and 1.33 for pion, rho and nucleon,
respectively.  In contrast, for the $1/V$-fits $\chi^2/\mathrm{dof}$
is 19, 8 and 4.  The asymptotic mass $\mpi$ ($\mrho$, $m_N$) obtained
from the fit deviates less than 1\% from $\Mpi$ ($\Mrho$, $M_N$) at
$L=24$ and lies thus well within the errors of $\Mpi$ ($\Mrho$,
$M_N$).

\section{CONCLUSIONS AND OUTLOOK}

We clearly see an exponential behaviour in our data for the light
hadrons, as opposed to the $1/L^3$-dependence reported earlier by
Fukugita \emph{et al.}. However, the $1/L^3$-fits are acceptable when
we shift the fit-range from $(12..24)$ towards smaller $L$, say
$(10..16)$.  Moreover, we have to allow for extra free parameters in
L\"uscher's formula in order to achieve a good description of our
data.  These parameters are $O(1000)$ in the case of $a_1$ and about 2
in the case of $a_2$ (except for the nucleon where $a_1 = O(100)$ and
$a_2 \approx 1$) and vary with the left fit boundary. This, together
with the $\Fpi L$-values mentioned above, suggests that we are in an
intermediate, sub-asymptotic regime were we merely see the onset of
the chiral behaviour predicted by L\"uscher's formula.

In a next step we will analyse our data at $\beta=5.6$, $\kappa=0.158$
and $\beta=5.32$, $\kappa=0.1665$ in order to check whether the
observed volume-dependence is reproduced there. We will then address
the question if and to what precision an extrapolation to the infinite
volume is possible.

\section*{ACKNOWLEDGEMENTS}

The gauge configurations were produced on the cluster computer ALiCE
at the University of Wuppertal and on the Quadrics/APE installation at
DESY/NIC Zeuthen. Further numerical calculations were carried out on
the Cray T3E at FZ/NIC J\"ulich. We thank the operating staffs in all
these places for their support. BO thanks S.~D\"urr for useful
discussions.

\end{document}